\documentclass[aps,pre,superscriptaddress,showpacs,twocolumn,longbibliography]{revtex4-1}
\usepackage{color}
\usepackage[usenames,dvipsnames]{xcolor}
\usepackage{graphicx}
\usepackage{color}
\usepackage{epsfig}
\usepackage{bm} 
\usepackage{amsmath}
\usepackage{amsfonts}
\usepackage{amssymb}

\usepackage{color}
\definecolor{myblue}{RGB}{65,105,225}
\definecolor{mygreen}{RGB}{34,139,34}
\definecolor{myorange}{RGB}{255,69,0}
\usepackage[colorlinks,linkcolor=myorange,urlcolor=myblue,citecolor=myblue]{hyperref}

\def\(({\left(}
\def\)){\right)}
\def\[[{\left[}
\def\]]{\right]}

\newcommand{\hD}{{\hat{D}}}
\newcommand{\hS}{{\hat{\sigma}}}

\newcommand{\be}{\begin{equation}}
\newcommand{\ee}{\end{equation}}
\newcommand{\beq}{\begin{equation}}
\newcommand{\eeq}{\end{equation}}
\newcommand{\ben}{\begin{eqnarray}}
\newcommand{\een}{\end{eqnarray}}

\newcommand{\la}{\langle}
\newcommand{\ra}{\rangle}

\newcommand{\vrr}{{\bf{r}}}
\newcommand{\vxi}{{\bm{\xi}}}
\newcommand{\vnabla}{{\bm{\nabla}}}
\newcommand{\vj}{{\bf{j}}}

\newcommand{\vE}{{\bf{E}}}
\newcommand{\vq}{{\bf{q}}}

\newcommand{\vlamb}{{\bm{\lambda}}}

\newcommand{\mA}{{\hat{\cal A}}}

\newcommand{\vchi}{{\bm{\varphi}}}

\begin{document}

\title{Structure of the optimal path to a fluctuation}

\author{N. Tiz{\'o}n-Escamilla}
\email[]{tizon@onsager.ugr.es}
\affiliation{Departamento de Electromagnetismo y F\'{\i}sica de la Materia, and Instituto Carlos I de F{\'\i}sica Te{\'o}rica y Computacional. Universidad de Granada. E-18071 Granada. Spain }

\author{P.I. Hurtado}
\email[]{phurtado@onsager.ugr.es}
\affiliation{Departamento de Electromagnetismo y F\'{\i}sica de la Materia, and Instituto Carlos I de F{\'\i}sica Te{\'o}rica y Computacional. Universidad de Granada. E-18071 Granada. Spain }

\author{P.L. Garrido}
\email[]{garrido@onsager.ugr.es}
\affiliation{Departamento de Electromagnetismo y F\'{\i}sica de la Materia, and Instituto Carlos I de F{\'\i}sica Te{\'o}rica y Computacional. Universidad de Granada. E-18071 Granada. Spain }

\date{\today}

\begin{abstract}
Macroscopic fluctuations have become an essential tool to understand physics far from equilibrium due to the link between their statistics and nonequilibrium ensembles. The optimal path leading to a fluctuation encodes key information on this problem, shedding light on e.g. the physics behind the enhanced probability of rare events out of equilibrium, the possibility of dynamic phase transitions and new symmetries. This makes the understanding of the properties of these optimal paths a central issue. Here we derive a fundamental relation which strongly constraints the architecture of these optimal paths for general $d$-dimensional nonequilibrium diffusive systems, and implies a non-trivial structure for the dominant current vector fields. Interestingly, this general relation (which encompasses and explains previous results) makes manifest the spatio-temporal non-locality of the current statistics and the associated optimal trajectories.
\end{abstract}


\maketitle

\section{Introduction}

Understanding the statistics of macroscopic fluctuations in nonequilibrium systems remains as a major challenge of theoretical physics. This interest is rooted in the prominent role that fluctuations play in equilibrium, where their statistics is directly linked to the relevant thermodynamic potentials via the Einstein formula \cite{einstein1910a,landau68a}. Similarly, it is nowadays expected that a deeper understanding of nonequilibrium fluctuations will pave the way to a sound definition of nonequilibrium potentials \cite{bertini15a,derrida07a,hurtado14a}, though we already know that these functions do typically have some striking features peculiar to nonequilibrium behavior (as e.g. non-local behavior leading to long-range correlations). Among all possible observables that can be defined, the currents of locally-conserved quantities play a key role as tokens of nonequilibrium physics, appearing in response to any driving mechanism (as e.g. a boundary gradient or external field) pushing the system out of equilibrium. In this way, the distribution of current fluctuations is a central object of investigation, with the associated current \emph{large deviation function} (LDF) \cite{touchette09a} acting as a marginal of the nonequilibrium analog of thermodynamic potential. 

In recent years, a macroscopic fluctuation theory (MFT) has been formulated \cite{bertini15a,bertini01a,bertini02a,bertini05a,bertini06a} to study dynamic fluctuations in systems far from equilibrium, starting from a mesoscopic description of the system of interest in terms of fluctuating hydrodynamics \cite{sasa08a,bertini12a,bertini13a,prados11a,prados12a,hurtado13a,jack15a,bodineau10a,krapivsky14a,meerson14a,bouchet16a}. Indeed MFT needs of only a few transport coefficients which can be easily determined in experiments or simulations. From this starting point, MFT offers detailed predictions for the large deviation functions of interest in terms of a complex spatio-temporal variational problem for the locally-conserved fields and the associated currents \cite{bertini15a}. As an interesting by-product, MFT also determines the optimal path to a fluctuation from the solution of the Euler-Lagrange equations for this variational problem. Understanding the properties and spatio-temporal structure of these optimal paths is of paramount importance, as they contain information on possible \emph{dynamic} phase transitions appearing at the fluctuating level \cite{bodineau05a,hurtado11a,perez-espigares13a,hurtado14a,zarfaty16a,vaikuntanathan14a,lam09a,chandler10a,shpielberg16a,baek16a,tizon-escamilla16a}, while their symmetry properties lead to new fluctuation theorems \cite{evans93a, gallavotti95a, gallavotti95b, kurchan98a, lebowitz99a, andrieux07a, gallavotti14a, hurtado11b, villavicencio14a, lacoste14a, gaspard13a, perez-espigares15a, kumar15a}.

The complexity of the MFT variational problem is such that most studies to date have focused on the current statistics of oversimplified one-dimensional ($1d$) transport models for which the MFT problem is somewhat simpler, specially when aided with the Additivity Principle \cite{bodineau04a, pilgram03a, jordan04a, bodineau06a, hurtado09c, hurtado10a, gorissen12a, gorissen12b, lazarescu15a}. Only very recently MFT has been used to understand current fluctuations in more realistic high-dimensional ($d>1$) systems \cite{hurtado11b, hurtado14a, akkermans13a, becker15a, perez-espigares16a, villavicencio16a, tizon-escamilla16a}, and these studies have unveiled a rich phenomenology which only appears for $d>1$, including hidden symmetries leading to new fluctuation theorems \cite{hurtado11b}, a weak generalization of the Additivity Principle \cite{perez-espigares16a}, and complex dynamic phase transitions associated to competing emergent orders and symmetry breaking phenomena \cite{tizon-escamilla16a}. Crucially, the richness found in $d>1$ stems in all cases from the relevance of structured optimal current fields at the fluctuating level, a common trait of all these new results \cite{perez-espigares16a,villavicencio16a,tizon-escamilla16a}. In this paper we show that structured optimal current fields are a \emph{fundamental requirement} of any high-dimensional fluctuating theory, rather than a mathematical accident. In particular, a simple calculation within MFT allows us to relate the Jacobian matrix of the reduced optimal current field (to be defined below) with the Hessian matrix of a response field which guarantees that the continuity equation (expressing the local conservation law) is fulfilled at all points of space and time. A natural analyticity requirement for this response field then leads to a strong condition on the reduced optimal current field: in brief, the optimal current vector field is bounded to exhibit non-trivial structure along the dominant direction in all its orthogonal components, and this structure is coupled to the optimal density field via the mobility transport coefficient. This coupling is explicitly non-local in space and time, a main feature of nonequilibrium physics. This result sheds new light and encompass all previous works on current fluctuations in $d>1$, opening the door to further developments in this field.

\begin{figure}
\vspace{-0.3cm}
\includegraphics[width=7.5cm]{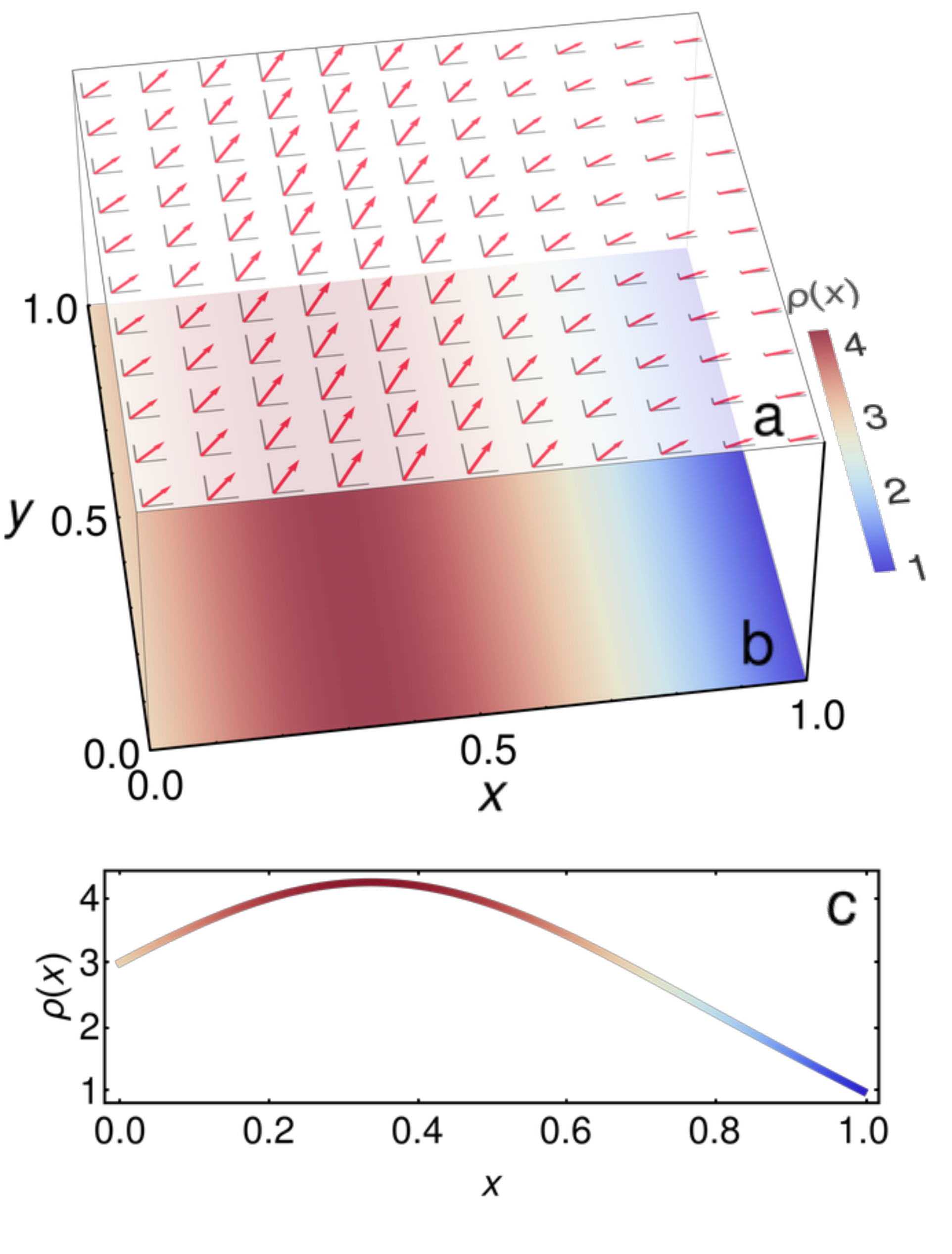}
\vspace{-0.2cm}
\caption{(Color online) Optimal solution for the current vector field (a) and the density field (b and c) associated to a given current fluctuation in the $2d$ Kipnis-Marchioro-Presutti model of heat transport in contact with two boundary thermal baths at temperatures $\rho(x=0)=\rho_0=3$ and $\rho(x=1)=\rho_1=1$ and no external field (in this model the locally-conserved \emph{density} field is the energy). Gray lines in (a) depict both local components of the optimal current vector field, while red arrows show the resultant vectors. Note the non-trivial structure of the $y$-component of the current field along the gradient $x$-direction, in stark contrast with the constant, structureless current $x$-component.
}
\label{fig1}
\end{figure}

To illustrate the meaning of the structure described above, we show in Fig.~\ref{fig1} both the optimal density $\rho(x)$ and current vector fields associated to a particular (rare) current fluctuation in a broadly studied driven diffusive system, the two-dimensional Kipnis-Marchioro-Presutti (KMP) model of heat transport in contact with two boundary thermal baths located at $x=0,1$ and no external field \cite{perez-espigares16a}. In this case, the dominant direction of structure formation corresponds to that of the temperature gradient, resulting in optimal density fields with structure only along the $x$-direction (Fig. \ref{fig1}.b-c). Consequently, the optimal current vector field exhibits a non-trivial structure in its $y$-component along the gradient $x$-direction, proportional to the local density field squared as dictated by the KMP mobility transport coefficient, which is simply $\sigma(\rho)=\rho^2$. This structure of the current $y$-component, which contrasts with the constant structureless $x$-component (Fig. \ref{fig1}.a), is the manifestation of a general theorem for driven diffusive systems that we prove next.

\section{Macroscopic Fluctuation Theory}

To be more precise, we focus now on a broad class of $d$-dimensional anisotropic driven diffusive systems characterized by a density field $\rho(\vrr,t)$, with $\vrr\in\Lambda\equiv[0,1]^d$ and $t\in[0,\tau]$, which represents any locally-conserved observable as e.g.~a density of particles, energy, charge, etc. This density field evolves in time according to the following fluctuating hydrodynamics equation \cite{bertini15a,derrida07a,hurtado14a}
\be
\partial_t \rho(\vrr,t) + \vnabla \cdot \left( -\hD(\rho) \vnabla \rho(\vrr,t) + \hS(\rho)\vE + \vxi(\vrr,t) \right) = 0 \, .
\label{langevin}
\ee
The field $\vchi(\vrr,t)\equiv -\hD(\rho) \vnabla \rho(\vrr,t) + \hS(\rho)\vE + \vxi(\vrr,t)$ acts as a fluctuating current, with $\vE$ an external driving field. In this way Eq.~(\ref{langevin}) is nothing but the continuity equation expressing the local conservation law. The deterministic part of the current field $\vchi(\vrr,t)$ is given by Fick's law under external driving, where $\hD(\rho)\equiv D(\rho)\mA$ and $\hS(\rho)=\sigma(\rho)\mA$ are the diffusivity and mobility matrices, respectively, and $\mA$ is a diagonal anisotropy matrix with components $\mA_{\alpha\beta}=a_\alpha\delta_{\alpha\beta}$, $\alpha,\beta\in [1,d]$, which we assume constant and independent of the local density. The vector field $\vxi(\vrr,t)$ is a Gaussian white noise term with zero average, $\la \vxi(\vrr,t)\ra=0$, and variance 
\be
\la \xi_\alpha(\vrr,t)\xi_\beta(\vrr',t') \ra=\frac{\sigma(\rho)}{L^d} a_\alpha\delta_{\alpha\beta}\delta(\vrr-\vrr')\delta(t-t') \, , \nonumber
\ee
with $L$ the system size in natural units. This (conserved) noise term accounts for the many fast microscopic degrees of freedom which are averaged out in the coarse-graining procedure leading to Eq.~(\ref{langevin}). Note that, at this mesoscopic level of description, the diffusion and mobility transport matrices fully characterize the dynamic and fluctuation properties of the model at hand. In general, systems described in this way are driven out of equilibrium by either (a) the action of the \emph{bulk} external field $\vE$, (b) a \emph{boundary} gradient imposed by appropriate boundary conditions on the density field (more on this below), or (c) possibly by the combined action of both (bulk + boundary) driving mechanisms. However, in the absence of driving, we expect the system to relax to equilibrium. In this case both transport coefficients cannot be independent, being related via a local Einstein relation $\hD(\rho)= \hS(\rho) f_0''(\rho)$, with $f_0(\rho)$ the \emph{equilibrium} free energy 
density of the system of interest and $'$ denoting differentitation with respect to the argument. Our results below can be however easily generalized to more general theories violating the previous condition.

Starting from the Fokker-Planck description of the Langevin equation (\ref{langevin}), and using a path integral representation, the probability of observing a particular \emph{trajectory} $\{\rho(\vrr,t),\vj(\vrr,t)\}_0^{\tau}$ of duration $\tau$ for the density and current fields can be written as \cite{bertini15a,derrida07a,hurtado14a}
\be
\text{P}\left(\{\rho,\vj\}_0^{\tau} \right) \sim \exp \Big( +L^d I_{\tau}\left[\rho,\vj \right] \Big) \, , 
\label{probpath} 
\ee
meaning that
\be
\lim_{L\to\infty} \frac{1}{L^d} \ln \text{P}\left(\{\rho,\vj\}_0^{\tau} \right) = I_{\tau}\left[\rho,\vj \right] \, . \nonumber
\label{asymp}
\ee
The action of Eq.~(\ref{probpath}) is
\be
I_{\tau}\left[\rho,\vj \right]=- \int_0^{\tau} dt \int_\Lambda d \vrr\, \frac{1}{2\sigma(\rho)}\bm{\mathcal J}(\vrr,t) \cdot \mA^{-1}\bm{\mathcal J}(\vrr,t) \,  , 
\label{action}
\ee
with the definition 
\be
\bm{\mathcal J}(\vrr,t)\equiv \vj+\hD(\rho) \vnabla\rho -\hS(\rho)\vE \, ,
\label{excesscurr}
\ee
and the additional constraint that the fields $\rho (\vrr,t)$ and $\vj(\vrr,t)$ must be coupled via the continuity equation at every point of space and time, see Eq.~(\ref{langevin}),
\be
\partial_t \rho (\vrr,t) + \vnabla\cdot \vj(\vrr,t) = 0 \, . 
\label{contappA}
\ee
For trajectories $\{\rho,\vj\}_0^{\tau}$ not obeying this continuity constraint or the appropriate boundary conditions (which depend on the particular problem at hand, see below), $I_{\tau}\left[\rho,\vj \right]\to-\infty$. Note that the field $\bm{\mathcal J}(\vrr,t)$ in Eq.~(\ref{action}) is nothing but the \emph{excess current}, i.e. the departure of the current vector field $\vj(\vrr,t)$ from its constitutive form $-\hD(\rho) \vnabla\rho + \hS(\rho)\vE$ associated to the coupled density field via Fick's law under external driving. Eq.~(\ref{action}) and the associated definitions constitute the fundamental formula of macroscopic fluctuation theory \cite{bertini15a}, from which many important and general results can be derived, valid arbitrarily far from equilibrium.

The probability $\text{P}_{\tau}(\vq)$ of observing a space- and time-averaged empirical current $\vq$, defined as
\be
\vq = \frac{1}{\tau}  \int_0^{\tau} dt \int_\Lambda d\vrr \, \vj (\vrr,t) \, ,
\label{currappA}
\ee
can be now obtained by summing up the probability of all trajectories $\{\rho,\vj\}_0^{\tau}$ compatible with the constraint (\ref{currappA}) on the empirical current and the continuity constraint (\ref{contappA}). Mathematically
\begin{eqnarray}\label{probcond1}
\displaystyle \text{P}_{\tau}\left(\vq \right)&=& \int {\cal D}\rho {\cal D}\vj \,\, \text{P}\left(\{\rho,\vj\}_0^{\tau} \right) \delta\left(\partial_{t} \rho+\vnabla \cdot \vj\right) \times \nonumber \\ 
&\times& \displaystyle\delta\left(\vq- \tau^{-1}  \int_0^{\tau} dt \int_\Lambda d\vrr\, \vj \right) \, , \nonumber
\end{eqnarray}
with the Dirac $\delta$-functionals guaranteeing the above contraints. We can now just use the Fourier-Laplace representation of these $\delta$-functionals, namely
\ben
\delta\left(\vq- \tau^{-1}  \int_0^{\tau} dt \int_\Lambda d\vrr\, \vj \right) = \int d\vlamb \, \text{e}^{- L^d \vlamb\cdot [\tau\vq -  \int_0^{\tau} dt \int_\Lambda d \vrr\, \vj(\vrr,t)]}\, , \label{delta1} \nonumber \\ 
\delta\left(\partial_{t} \rho+\vnabla \cdot \vj\right) = \int {\cal D} \psi \, \text{e}^{-L^d \int_0^{\tau} dt \int_\Lambda d \vrr\, \psi(\vrr,t)(\partial_{t} \rho+\vnabla \cdot \vj)} \, , \label{delta2}  \nonumber
\een
and the large deviation principle Eq.~(\ref{probpath}) to arrive at
\be
\text{P}_\tau \left(\vq \right)\sim \int {\cal D}\rho\, {\cal D}\vj\,  {\cal D}\psi\, d\vlamb \, \exp \left( + L^d {\cal I}_\tau\left[\rho,\vj,\psi,\vlamb \right] \right) \nonumber
\ee
where the modified action reads
 \begin{eqnarray}
\displaystyle {\cal I}_\tau \left[\rho,\vj,\psi,\vlamb \right]&=&- \int_0^{\tau} dt \int_\Lambda d \vrr \Big[\frac{1}{2\sigma(\rho)}\bm{\mathcal J}\cdot \mA^{-1} \bm{\mathcal J} + \nonumber \label{modaction} \\
&+&\psi(\vrr,t)\left(\partial_{t}\rho+\vnabla \cdot j\right)+\vlamb\cdot [\vq-\vj(\vrr,t)]\Big] \nonumber \, .
 \end{eqnarray}
For long times and large system sizes, the probability of observing an empirical current $\vq$ peaks around the average current $\la \vq \ra$ as $\text{P}_{\tau}(\vq)\sim \exp[+\tau L^d G(\vq)]$, and the concentration rate $G(\vq)$ defines the current large deviation function. This is a measure of the exponential rate at which the likelihood of observing a current $\vq\ne\la\vq\ra$ decays as $\tau$ and $L$ increase (note that, consequently, $G(\la\vq\ra)=0$). In this limit, the current LDF can be written as 
\be
G(\vq) = \lim_{\tau \to \infty}  \frac{1}{\tau} \max_{\{\rho,\vj,\psi,\vlamb\}_0^\tau} {\cal  I}_\tau \left[\rho,\vj,\psi,\vlamb \right]  \, .
\label{LDFappA}
\ee

\section{Structure of the optimal path}

The set $\left(\rho_\vq,\vj_\vq,\psi_\vq,\vlamb_\vq\right)$ of optimal fields which solve this variational problem define \emph{the most probable path} leading to a current fluctuation $\vq$. Equations for these optimal fields can be derived now by functional differentiation of the above modified action. In particular, by varying over the density field, $\rho(\vrr,t)\to \rho(\vrr,t) + \delta\rho(\vrr,t)$, we arrive at the following partial differential equation 
\be
\partial_t\psi_\vq= H(\rho_\vq)-\frac{\sigma'_\vq}{2\sigma_\vq^2}\vj_\vq\cdot\mA^{-1}\vj_\vq+\frac{\sigma_\vq}{2}\vE\cdot\mA\vE \, , \nonumber
\label{optdens}
\ee
where we have defined
\be
H(\rho_\vq) \equiv - \left[\vnabla\left(\frac{D_\vq^2}{2\sigma_\vq}\right)+\frac{D_\vq^2}{\sigma_\vq}\vnabla\right]\cdot\mA\vnabla\rho_\vq \, , \nonumber
\label{Fdef}
\ee
with $D_\vq\equiv D(\rho_\vq)$ and $\sigma_\vq\equiv \sigma(\rho_\vq)$. Another equation is obtained by varying over the current field, $\vj\to \vj + \delta \vj$, leading to
\be
\bm{\mathcal J}_\vq=\hS_\vq\left(\vlamb_\vq+\vnabla\psi_\vq\right) \,
\label{optcurr}
\ee
where $\bm{\mathcal J}_\vq = \vj_\vq+\hD_\vq \vnabla\rho_\vq -\hS_\vq\vE$ is the optimal excess current, see Eq.~(\ref{excesscurr}). Finally, variations over $\psi$ and $\vlamb$ lead respectively to the constraints (\ref{contappA}) and (\ref{currappA}) for the optimal density and current fields, $\rho_\vq(\vrr,t)$ and $\vj_\vq(\vrr,t)$.

Before continuing, we can now gain some insight on the physical interpretation of $\vlamb_\vq$ and $\psi_\vq$ by using the local Einstein formula $\hD_\vq= \hS_\vq f_0''(\rho_\vq)$ to write Fick's law under external driving as $-\hD_\vq \vnabla\rho_\vq + \hS_\vq\vE= \hS_\vq [\vE- \vnabla (\delta {\cal F}_0/\delta \rho_\vq)]$, where ${\cal F}_0(\rho)=\int_\Lambda d\vrr \, f_0(\rho)$ is the equilibrium free energy functional of the system of interest. Using this in Eq.~(\ref{optcurr}) we find that
\be
\vj_\vq= \hS_\vq \left[ (\vE+\vlamb_\vq) - \vnabla \left( \frac{\delta {\cal F}_0}{\delta\rho_\vq} - \psi_\vq\right)  \right] \, . 
\label{optcurr2}
\ee
In this way, $\vlamb_\vq$ and $\psi_\vq(\vrr,t)$ can be interpreted respectively as the additional bulk field and boundary driving (i.e. chemical potential) necessary to obtain the current field $\vj_\vq(\vrr,t)$ within Fick's law under external driving. Alternatively, note also that $\psi_\vq$ is nothing but the (optimal) Legendre multiplier associated to continuity equation, Eq. \ref{contappA}, and as such it is intimately related to the noise field. Indeed, the field $\psi$ selects those noise realizations compatible with Fick's law and the local conservation law (this can be better seen in the Hamiltonian formulation of the problem \cite{baek16a,martin16a} where $\psi$ plays the role of the conjugate moment to the density).

Eq.~(\ref{optcurr}), or equivalently Eq.~(\ref{optcurr2}), sets strong conditions on the structure of the optimal current field. In particular, if we define now the reduced (optimal) excess current $\bm{\chi}_\vq(\vrr,t) \equiv \hS_\vq^{-1}\bm{\mathcal J}_\vq(\vrr,t)$ and take its Jacobian matrix $\vnabla\bm{\chi}_\vq$, with components $(\vnabla\bm{\chi}_\vq)_{\alpha\beta}=\partial_\alpha {\chi}_{\vq,\beta}$, we have from Eq.~(\ref{optcurr}) that $\vnabla\bm{\chi}_\vq = \vnabla \vnabla \psi_\vq$, or equivalently
\be
\partial_\alpha {\chi}_{\vq,\beta} = \partial_\alpha \partial_\beta \psi_\vq \, .
\label{cond1}
\ee
In words, the Jacobian matrix of the reduced (optimal) excess current $\bm{\chi}_\vq$ corresponds to the Hessian of the response field $\psi_\vq$ associated to the continuity equation (\ref{contappA}). This observation thus leads to the following strong result:
\\

\emph{\bf Theorem}: Let the response function $\psi_\vq:\Lambda^d\times[0,\tau] \to \mathbb{R}$ be a $C^2$-class function of spatial coordinates, i.e. a function twice continuously differentiable in its spatial domain. Then 
\be
\partial_\beta\left(\frac{\displaystyle j_{\alpha,\vq}}{\displaystyle a_\alpha \sigma_\vq}\right) = \partial_\alpha\left(\frac{\displaystyle j_{\beta,\vq}}{\displaystyle a_\beta\sigma_\vq}\right) \quad \forall \, (\vrr,t) \in\Lambda^d\times [0,\tau] \, .
\label{condition}
\ee
\indent \emph{\bf Proof}: Schwarz's theorem \cite{arfken11a} states that if a function $\psi_\vq$ has continuous second partial derivatives at any given spatial point in $\Lambda^d$ then its Hessian matrix is symmetric at this point, $\partial_\alpha \partial_\beta \psi_\vq=\partial_\beta \partial_\alpha \psi_\vq$. This immediately implies, via Eq.~(\ref{cond1}), that the Jacobian of the reduced (optimal) excess current is itself a symmetric matrix, i.e. $\partial_\alpha {\chi}_{\vq,\beta}=\partial_\beta {\chi}_{\vq,\alpha}$ $\forall \alpha,\beta\in[1,d]$. From this symmetry, and using the definitions of $\bm{\chi}_\vq$ and $\bm{\mathcal J}_\vq$ above, and the relation $\partial_\alpha (D_\vq/\sigma_\vq)\partial_\beta\rho_\vq = \partial_\beta (D_\vq/\sigma_\vq)\partial_\alpha\rho_\vq = (D_\vq/\sigma_\vq)' \partial_\alpha\rho_\vq \partial_\beta\rho_\vq$, we immediately arrive at the fundamental relation (\ref{condition}). Note that the $C^2$-differentiability of the response function $\psi_\vq$ is a \emph{natural} requirement for most physical solutions to this variational problem, though we cannot discard the possible existence of singular, non-differentiable solutions for $\psi_\vq$ which would violate (\ref{condition}) at singular points. Note also that a weaker condition for $\psi_\vq$ which nevertheless suffices to ensure the symmetry of its Hessian matrix is that all partial derivatives are themselves differentiable.

To better understand the tight constraints that Eq.~(\ref{condition}) imposes on the optimal current fields, it is important to realize that in all high-dimensional problems studied in literature up to now the dominant paths responsible of a current fluctuation, corresponding to the global extrema of the action ${\cal I}_\tau$ in Eq.~(\ref{LDFappA}), always exhibit structure (if any) along a \emph{principal direction}, that we denote here as $x_\parallel$ \cite{hurtado14a,perez-espigares16a,tizon-escamilla16a,akkermans13a,becker15a,villavicencio16a}. This means in particular that $\rho_\vq(\vrr,t)=\rho_\vq(x_\parallel,t)$ and $\vj_\vq(\vrr,t)=\vj_\vq(x_\parallel,t)$. Examples include open systems subject to a boundary gradient, which develop structure along the gradient direction (irrespective of the external field) \cite{perez-espigares16a}, see e.g. Fig. \ref{fig1} above; or closed driven diffusive systems with periodic boundary conditions, for which different dynamic phase transitions appear to current regimes characterized by traveling waves with structure along one of the principal axes of the system of interest \cite{tizon-escamilla16a}. In all these cases, condition (\ref{condition}) leads to
\be
\partial_\parallel\left(\frac{\displaystyle j_{\beta,\vq}}{\displaystyle a_\beta \sigma_\vq}\right) = 0 \qquad \forall \, \beta\ne \parallel \, , \nonumber
\label{condition2}
\ee
which immediately implies that $ j_{\beta,\vq}(x_\parallel,t) = k_\beta \sigma[\rho_\vq(x_\parallel,t)]$ $\forall \beta\ne \parallel$, with $k_\beta$ a direction-dependent constant which follows from the constraint (\ref{currappA}) on the empirical current $\vq$. Therefore we arrive at
\be
j_{\beta,\vq}(x_\parallel,t) = q_\beta \frac{\displaystyle  \tau \sigma[\rho_\vq(x_\parallel,t)]}{\displaystyle  \int_0^\tau ds \int_0^1 dy \, \sigma[\rho_\vq(y,s)]} \qquad \forall \, \beta\ne \parallel \, .
\label{condition3}
\ee
In this way the relation between the Jacobian matrix for $\bm{\chi}_\vq$ and the Hessian matrix of the response field $\psi_\vq$, together with a natural analyticity condition for the latter, force the optimal current vector field $\vj_\vq$ to exhibit \emph{non-trivial structure} along the dominant direction $\parallel$ in all its orthogonal components $\beta\ne\parallel$, and this structure is coupled to the optimal density field $\rho_\vq$ via the mobility transport coefficient $\sigma(\rho_\vq)$.  Interestingly, this result makes manifest the \emph{spatio-temporal nonlocality} of the current LDF (\ref{LDFappA}) and the associated optimal trajectories, as the optimal current field at a given point of space and time depends explicitly on the space-time integral of the mobility of the optimal density field, see the denominator in Eq.~(\ref{condition3}). Note also that conditions (\ref{condition}) and (\ref{condition3}) become empty for $d=1$, where structureless optimal current fields are still possible \cite{bodineau04a, hurtado09c, hurtado10a}, evidencing the richness of the fluctuation landscape for $d>1$ driven diffusive systems when compared with their one-dimensional counterparts.

\section{Connection with previous results}

We next explore how previous results on current fluctuations for both open and closed $d>1$ driven diffusive systems fit into the above general result. First we consider the case of open systems under an external gradient along an arbitrary direction $x_\parallel$. For that we fix the boundary densities to $\rho(\vrr,t)\vert_{x_\parallel=0,1} = \rho_{0,1}$, which drive the system out of equilibrium as soon as $\rho_0\ne\rho_1$, setting periodic boundary conditions for all other directions of space. This class of systems has been broadly studied during the last years, finding that a simplifying conjecture within MFT known as (weak) Additivity Principle (AP) \cite{perez-espigares16a} allows to solve the problem of current statistics both for $d=1$ \cite{bodineau04a,hurtado09c,hurtado10a,hurtado14a,gorissen12a} and $d>1$ \cite{saito11a, akkermans13a, becker15a, villavicencio16a, perez-espigares16a}. The AP, which offers explicit expressions for the current LDF and the optimal paths supporting a given fluctuation, establishes that the most probable trajectory to a current fluctuation is time-independent (apart from some initial and final transients of negligible weight for the current LDF). In this case $\rho_\vq=\rho_\vq(x_\parallel)$ and $\vj_\vq=\vj_\vq(x_\parallel)$, so the continuity constraint Eq.~(\ref{contappA}) implies a divergence-free optimal current field, $\vnabla\cdot\vj_\vq(x_\parallel) = \partial_\parallel j_{\parallel,\vq}(x_\parallel) = 0$. These observations, together with our general condition (\ref{condition3}) and the constraint (\ref{currappA}) on the empirical current $\vq$, lead to an optimal current field $\vj_\vq(x_\parallel)=(q_\parallel,\vj_{\perp,\vq})$ with
\be
\vj_{\perp,\vq}(x_\parallel) = \vq_\perp \frac{\displaystyle \sigma[\rho_\vq(x_\parallel)]}{\displaystyle  \int_0^1 dy \, \sigma[\rho_\vq(y)]}  \, , \nonumber
\label{condition3AP}
\ee
and where we have decomposed $\vq = (q_\parallel,\vq_\perp)$ along the gradient ($\parallel$) and all other, $(d-1)$ directions ($\perp$). This corresponds exactly to the result obtained previously from the weak Additivity Principle as applied to $d$-dimensional driven diffusive systems \cite{perez-espigares16a}, starting from a variational problem for general but divergence-free current fields with structure along one dominant direction. Our general theorem (\ref{condition}) allows now to understand this structure as a direct consequence of the symmetry of the Jacobian matrix associated to the reduced excess current. Note that this result is not compatible with the straightforward extension to $d>1$ of the $1d$-system solution (which considers the optimal current field to be constant \cite{bodineau04a,bodineau06a,hurtado09c,hurtado10a}).

To end this paper, we consider current fluctuations in \emph{closed} $d$-dimensional anisotropic driven diffusive systems under an external field $\vE$ \cite{tizon-escamilla16a}. For that we set periodic boundary conditions along all directions of space. Due to the system periodicity, the total \emph{mass} is conserved so $\rho_0=\int_\Lambda \rho_\vq(\vrr,t) d\vrr$ is constant in time, a further constraint that has to be taken care of in the MFT variational problem. A detailed analysis of the resulting MFT equations shows \cite{bodineau05a,hurtado11a,hurtado14a,perez-espigares13a,tizon-escamilla16a} that a $2^\text{nd}$-order \emph{dynamic phase transition} (DPT) appears at a given critical current for this broad family of systems between a homogeneous fluctuation phase with Gaussian current statistics and constant, structureless optimal fields, $\rho_\vq(\vrr,t)=\rho_0$ and $\vj_\vq(\vrr,t)=\vq$, and a symmetry-broken non-Gaussian phase characterized by the emergence of coherent traveling waves with 
structure along a dominant direction, $\rho_\vq(\vrr,t)=\omega_\vq(x_\parallel-vt)$ and $\vj_\vq(\vrr,t)=\vj_\vq(x_\parallel-vt)$, with $v$ some velocity \cite{tizon-escamilla16a}. Interestingly, for mild or no anisotropy, different traveling-wave phases appear depending on the current separated by lines of $1^\text{st}$-order DPTs, a degeneracy which disappears beyond a critical anisotropy. This richness of the fluctuation phase diagram stems again from the relevance of structured current fields at the fluctuating level, a seemingly mathematical accident which takes full significance at the light of our general result (\ref{condition}). In particular, the continuity equation $\partial_t\rho_\vq + \vnabla\cdot\vj_\vq=0$ applied to the $1d$ traveling-wave structure implies that $\partial_\parallel [j_{\parallel,\vq}(z_\parallel) - v\omega(z_\parallel)]=0$, where we have defined $z_\parallel\equiv x_\parallel - vt$, and this together with the constraint (\ref{currappA}) on the empirical current leads to $j_{\parallel,\vq}(z_{\parallel})=q_{\parallel} - v[\rho_0-\omega_\vq(z_{\parallel})]$. On the other hand, all orthogonal current components follow directly from our theorem above as $\vj_{\perp,\vq}(z_\parallel) = \vq_\perp \sigma[\omega_\vq(z_\parallel)]/ \int_0^1 dy \, \sigma[\omega_\vq(y)]$. This result, which is markedly different from the traveling-wave structure found in $1d$ models \cite{bodineau05a,hurtado11b,perez-espigares13a}, has been recently derived within MFT after a careful analysis of the local stability of the homogeneous, Gaussian current phase against small but otherwise arbitrary spatio-temporal perturbations \cite{tizon-escamilla16a}. However its understanding as a direct consequence of the general condition (\ref{condition}) sheds new light onto this problem.

\section{Conclusions}

In summary, we have derived a fundamental relation which strongly constraints the structure of the optimal path sustaining a given current fluctuation. In particular, when a principal direction exists, the optimal current vector field is bounded to exhibit non-trivial structure along this dominant direction in all its orthogonal components, a structure coupled to the optimal density field via the mobility transport coefficient. This has been done by relating within macroscopic fluctuation theory the Jacobian matrix of the reduced optimal current field with the Hessian matrix of the response field associated to the continuity equation, and requiring analyticity for the latter. In this sense, we prove here that the structured optimal current fields predicted and observed by a number of recent works \cite{perez-espigares16a,villavicencio16a,tizon-escamilla16a} is indeed a fundamental requirement of any high-dimensional fluctuating theory, rather than a mathematical accident. Remarkably, our result also makes manifest the non-locality in space and time of the current large deviation function and the associated optimal trajectories. This result hence serves as a starting point in the study of fluctuations in complex $d$-dimensional systems, constraining the form of the optimal paths and thus aiding in the formulation of simplifying hypotheses to solve these complex variational problems in nonequilibrium statistical physics.

\begin{acknowledgments}
We thank C. P\'erez-Espigares for useful discussions. Financial support from Spanish project FIS2013-43201-P (Ministerio de Econom\'ia y Competitividad) and FPU13/05633 is acknowledged.
\end{acknowledgments}


\begin{thebibliography}{59}%
\makeatletter
\providecommand \@ifxundefined [1]{%
 \@ifx{#1\undefined}
}%
\providecommand \@ifnum [1]{%
 \ifnum #1\expandafter \@firstoftwo
 \else \expandafter \@secondoftwo
 \fi
}%
\providecommand \@ifx [1]{%
 \ifx #1\expandafter \@firstoftwo
 \else \expandafter \@secondoftwo
 \fi
}%
\providecommand \natexlab [1]{#1}%
\providecommand \enquote  [1]{``#1''}%
\providecommand \bibnamefont  [1]{#1}%
\providecommand \bibfnamefont [1]{#1}%
\providecommand \citenamefont [1]{#1}%
\providecommand \href@noop [0]{\@secondoftwo}%
\providecommand \href [0]{\begingroup \@sanitize@url \@href}%
\providecommand \@href[1]{\@@startlink{#1}\@@href}%
\providecommand \@@href[1]{\endgroup#1\@@endlink}%
\providecommand \@sanitize@url [0]{\catcode `\\12\catcode `\$12\catcode
  `\&12\catcode `\#12\catcode `\^12\catcode `\_12\catcode `\%12\relax}%
\providecommand \@@startlink[1]{}%
\providecommand \@@endlink[0]{}%
\providecommand \url  [0]{\begingroup\@sanitize@url \@url }%
\providecommand \@url [1]{\endgroup\@href {#1}{\urlprefix }}%
\providecommand \urlprefix  [0]{URL }%
\providecommand \Eprint [0]{\href }%
\providecommand \doibase [0]{http://dx.doi.org/}%
\providecommand \selectlanguage [0]{\@gobble}%
\providecommand \bibinfo  [0]{\@secondoftwo}%
\providecommand \bibfield  [0]{\@secondoftwo}%
\providecommand \translation [1]{[#1]}%
\providecommand \BibitemOpen [0]{}%
\providecommand \bibitemStop [0]{}%
\providecommand \bibitemNoStop [0]{.\EOS\space}%
\providecommand \EOS [0]{\spacefactor3000\relax}%
\providecommand \BibitemShut  [1]{\csname bibitem#1\endcsname}%
\let\auto@bib@innerbib\@empty
\bibitem [{\citenamefont {Einstein}(1910)}]{einstein1910a}%
  \BibitemOpen
  \bibfield  {author} {\bibinfo {author} {\bibfnamefont {A}~\bibnamefont
  {Einstein}},\ }\bibfield  {title} {\enquote {\bibinfo {title} {Theorie der
  opaleszenz von homogenen fl{\"u}ssigkeiten und fl{\"u}ssigkeitsgemischen in
  der n{\"a}he des kritischen zustandes},}\ }\href@noop {} {\bibfield
  {journal} {\bibinfo  {journal} {Ann. Phys. (Berlin)}\ }\textbf {\bibinfo
  {volume} {33}},\ \bibinfo {pages} {1275} (\bibinfo {year}
  {1910})}\BibitemShut {NoStop}%
\bibitem [{\citenamefont {Landau}\ and\ \citenamefont
  {Lifshitz}(1968)}]{landau68a}%
  \BibitemOpen
  \bibfield  {author} {\bibinfo {author} {\bibfnamefont {L.}~\bibnamefont
  {Landau}}\ and\ \bibinfo {author} {\bibfnamefont {E.}~\bibnamefont
  {Lifshitz}},\ }\href@noop {} {\emph {\bibinfo {title} {Course of theoretical
  physics. Vol. 5: Statistical physics}}}\ (\bibinfo  {publisher} {Pergamon
  Press},\ \bibinfo {year} {1968})\BibitemShut {NoStop}%
\bibitem [{\citenamefont {Bertini}\ \emph {et~al.}(2015)\citenamefont
  {Bertini}, \citenamefont {Sole}, \citenamefont {Gabrielli}, \citenamefont
  {Jona-Lasinio},\ and\ \citenamefont {Landim}}]{bertini15a}%
  \BibitemOpen
  \bibfield  {author} {\bibinfo {author} {\bibfnamefont {L.}~\bibnamefont
  {Bertini}}, \bibinfo {author} {\bibfnamefont {A.~De}\ \bibnamefont {Sole}},
  \bibinfo {author} {\bibfnamefont {D.}~\bibnamefont {Gabrielli}}, \bibinfo
  {author} {\bibfnamefont {G.}~\bibnamefont {Jona-Lasinio}}, \ and\ \bibinfo
  {author} {\bibfnamefont {C.}~\bibnamefont {Landim}},\ }\bibfield  {title}
  {\enquote {\bibinfo {title} {Macroscopic fluctuation theory},}\ }\href
  {http://journals.aps.org/rmp/abstract/10.1103/RevModPhys.87.593} {\bibfield
  {journal} {\bibinfo  {journal} {Rev. Mod. Phys.}\ }\textbf {\bibinfo {volume}
  {87}},\ \bibinfo {pages} {593--636} (\bibinfo {year} {2015})}\BibitemShut
  {NoStop}%
\bibitem [{\citenamefont {Derrida}()}]{derrida07a}%
  \BibitemOpen
  \bibfield  {author} {\bibinfo {author} {\bibfnamefont {B.}~\bibnamefont
  {Derrida}},\ }\bibfield  {title} {\enquote {\bibinfo {title} {Non-equilibrium
  steady states: fluctuations and large deviations of the density and of the
  current},}\ }\href
  {http://iopscience.iop.org/article/10.1088/1742-5468/2007/07/P07023}
  {\bibinfo  {journal} {J. Stat. Mech.: Theory Exp. (2007) P07023}}\BibitemShut {NoStop}%
\bibitem [{\citenamefont {Hurtado}\ \emph {et~al.}(2014)\citenamefont
  {Hurtado}, \citenamefont {Espigares}, \citenamefont {del Pozo},\ and\
  \citenamefont {Garrido}}]{hurtado14a}%
  \BibitemOpen
\bibfield  {journal} {  }\bibfield  {author} {\bibinfo {author} {\bibfnamefont
  {P.~I.}\ \bibnamefont {Hurtado}}, \bibinfo {author} {\bibfnamefont {C.~P.}\
  \bibnamefont {Espigares}}, \bibinfo {author} {\bibfnamefont {J.~J.}\
  \bibnamefont {del Pozo}}, \ and\ \bibinfo {author} {\bibfnamefont {P.~L.}\
  \bibnamefont {Garrido}},\ }\bibfield  {title} {\enquote {\bibinfo {title}
  {Thermodynamics of currents in nonequilibrium diffusive systems: theory and
  simulation},}\ }\href
  {http://link.springer.com/article/10.1007/s10955-013-0894-6} {\bibfield
  {journal} {\bibinfo  {journal} {J. Stat. Phys.}\ }\textbf {\bibinfo {volume}
  {154}},\ \bibinfo {pages} {214--264} (\bibinfo {year} {2014})}\BibitemShut
  {NoStop}%
\bibitem [{\citenamefont {Touchette}(2009)}]{touchette09a}%
  \BibitemOpen
  \bibfield  {author} {\bibinfo {author} {\bibfnamefont {H.}~\bibnamefont
  {Touchette}},\ }\bibfield  {title} {\enquote {\bibinfo {title} {The large
  deviation approach to statistical mechanics},}\ }\href
  {http://dx.doi.org/10.1016/j.physrep.2009.05.002} {\bibfield  {journal}
  {\bibinfo  {journal} {Phys. Rep.}\ }\textbf {\bibinfo {volume} {478}},\
  \bibinfo {pages} {1--69} (\bibinfo {year} {2009})}\BibitemShut {NoStop}%
\bibitem [{\citenamefont {Bertini}\ \emph {et~al.}(2001)\citenamefont
  {Bertini}, \citenamefont {Sole}, \citenamefont {Gabrielli}, \citenamefont
  {Jona-Lasinio},\ and\ \citenamefont {Landim}}]{bertini01a}%
  \BibitemOpen
  \bibfield  {author} {\bibinfo {author} {\bibfnamefont {L.}~\bibnamefont
  {Bertini}}, \bibinfo {author} {\bibfnamefont {A.~De}\ \bibnamefont {Sole}},
  \bibinfo {author} {\bibfnamefont {D.}~\bibnamefont {Gabrielli}}, \bibinfo
  {author} {\bibfnamefont {G.}~\bibnamefont {Jona-Lasinio}}, \ and\ \bibinfo
  {author} {\bibfnamefont {C.}~\bibnamefont {Landim}},\ }\bibfield  {title}
  {\enquote {\bibinfo {title} {Fluctuations in stationary nonequilibrium states
  of irreversible processes},}\ }\href
  {http://dx.doi.org/10.1103/PhysRevLett.87.040601} {\bibfield  {journal}
  {\bibinfo  {journal} {Phys. Rev. Lett.}\ }\textbf {\bibinfo {volume} {87}},\
  \bibinfo {pages} {040601} (\bibinfo {year} {2001})}\BibitemShut {NoStop}%
\bibitem [{\citenamefont {Bertini}\ \emph {et~al.}(2002)\citenamefont
  {Bertini}, \citenamefont {Sole}, \citenamefont {Gabrielli}, \citenamefont
  {Jona-Lasinio},\ and\ \citenamefont {Landim}}]{bertini02a}%
  \BibitemOpen
  \bibfield  {author} {\bibinfo {author} {\bibfnamefont {L.}~\bibnamefont
  {Bertini}}, \bibinfo {author} {\bibfnamefont {A.~De}\ \bibnamefont {Sole}},
  \bibinfo {author} {\bibfnamefont {D.}~\bibnamefont {Gabrielli}}, \bibinfo
  {author} {\bibfnamefont {G.}~\bibnamefont {Jona-Lasinio}}, \ and\ \bibinfo
  {author} {\bibfnamefont {C.}~\bibnamefont {Landim}},\ }\bibfield  {title}
  {\enquote {\bibinfo {title} {Macroscopic fluctuation theory for stationary
  non-equilibrium states},}\ }\href
  {http://link.springer.com/article/10.1023/A:1014525911391} {\bibfield
  {journal} {\bibinfo  {journal} {J. Stat. Phys.}\ }\textbf {\bibinfo {volume}
  {107}},\ \bibinfo {pages} {635--675} (\bibinfo {year} {2002})}\BibitemShut
  {NoStop}%
\bibitem [{\citenamefont {Bertini}\ \emph {et~al.}(2005)\citenamefont
  {Bertini}, \citenamefont {Sole}, \citenamefont {Gabrielli}, \citenamefont
  {Jona-Lasinio},\ and\ \citenamefont {Landim}}]{bertini05a}%
  \BibitemOpen
  \bibfield  {author} {\bibinfo {author} {\bibfnamefont {L.}~\bibnamefont
  {Bertini}}, \bibinfo {author} {\bibfnamefont {A.~De}\ \bibnamefont {Sole}},
  \bibinfo {author} {\bibfnamefont {D.}~\bibnamefont {Gabrielli}}, \bibinfo
  {author} {\bibfnamefont {G.}~\bibnamefont {Jona-Lasinio}}, \ and\ \bibinfo
  {author} {\bibfnamefont {C.}~\bibnamefont {Landim}},\ }\bibfield  {title}
  {\enquote {\bibinfo {title} {Current fluctuations in stochastic lattice
  gases},}\ }\href
  {http://journals.aps.org/prl/abstract/10.1103/PhysRevLett.94.030601}
  {\bibfield  {journal} {\bibinfo  {journal} {Phys. Rev. Lett.}\ }\textbf
  {\bibinfo {volume} {94}},\ \bibinfo {pages} {030601} (\bibinfo {year}
  {2005})}\BibitemShut {NoStop}%
\bibitem [{\citenamefont {Bertini}\ \emph {et~al.}(2006)\citenamefont
  {Bertini}, \citenamefont {Sole}, \citenamefont {Gabrielli}, \citenamefont
  {Jona-Lasinio},\ and\ \citenamefont {Landim}}]{bertini06a}%
  \BibitemOpen
  \bibfield  {author} {\bibinfo {author} {\bibfnamefont {L.}~\bibnamefont
  {Bertini}}, \bibinfo {author} {\bibfnamefont {A.~De}\ \bibnamefont {Sole}},
  \bibinfo {author} {\bibfnamefont {D.}~\bibnamefont {Gabrielli}}, \bibinfo
  {author} {\bibfnamefont {G.}~\bibnamefont {Jona-Lasinio}}, \ and\ \bibinfo
  {author} {\bibfnamefont {C.}~\bibnamefont {Landim}},\ }\bibfield  {title}
  {\enquote {\bibinfo {title} {Nonequilibrium current fluctuations in
  stochastic lattice gases},}\ }\href
  {http://link.springer.com/article/10.1007/s10955-006-9056-4} {\bibfield
  {journal} {\bibinfo  {journal} {J. Stat. Phys.}\ }\textbf {\bibinfo {volume}
  {123}},\ \bibinfo {pages} {237--276} (\bibinfo {year} {2006})}\BibitemShut
  {NoStop}%
\bibitem [{\citenamefont {Sasa}(2008)}]{sasa08a}%
  \BibitemOpen
  \bibfield  {author} {\bibinfo {author} {\bibfnamefont {S.}~\bibnamefont
  {Sasa}},\ }\bibfield  {title} {\enquote {\bibinfo {title} {A perturbation
  theory for large deviation functionals in fluctuating hydrodynamics},}\
  }\href {http://iopscience.iop.org/article/10.1088/1751-8113/41/4/045006}
  {\bibfield  {journal} {\bibinfo  {journal} {J. Phys. A}\ }\textbf {\bibinfo
  {volume} {41}},\ \bibinfo {pages} {045006} (\bibinfo {year}
  {2008})}\BibitemShut {NoStop}%
\bibitem [{\citenamefont {Bertini}\ \emph {et~al.}(2012)\citenamefont
  {Bertini}, \citenamefont {Gabrielli}, \citenamefont {Jona-Lasinio},\ and\
  \citenamefont {Landim}}]{bertini12a}%
  \BibitemOpen
  \bibfield  {author} {\bibinfo {author} {\bibfnamefont {L.}~\bibnamefont
  {Bertini}}, \bibinfo {author} {\bibfnamefont {D.}~\bibnamefont {Gabrielli}},
  \bibinfo {author} {\bibfnamefont {G.}~\bibnamefont {Jona-Lasinio}}, \ and\
  \bibinfo {author} {\bibfnamefont {C.}~\bibnamefont {Landim}},\ }\bibfield
  {title} {\enquote {\bibinfo {title} {Thermodynamic transformations of
  nonequilibrium states},}\ }\href
  {http://link.springer.com/article/10.1007/s10955-012-0624-5} {\bibfield
  {journal} {\bibinfo  {journal} {J. Stat. Phys.}\ }\textbf {\bibinfo {volume}
  {149}},\ \bibinfo {pages} {773--802} (\bibinfo {year} {2012})}\BibitemShut
  {NoStop}%
\bibitem [{\citenamefont {Bertini}\ \emph {et~al.}(2013)\citenamefont
  {Bertini}, \citenamefont {Gabrielli}, \citenamefont {Jona-Lasinio},\ and\
  \citenamefont {Landim}}]{bertini13a}%
  \BibitemOpen
  \bibfield  {author} {\bibinfo {author} {\bibfnamefont {L.}~\bibnamefont
  {Bertini}}, \bibinfo {author} {\bibfnamefont {D.}~\bibnamefont {Gabrielli}},
  \bibinfo {author} {\bibfnamefont {G.}~\bibnamefont {Jona-Lasinio}}, \ and\
  \bibinfo {author} {\bibfnamefont {C.}~\bibnamefont {Landim}},\ }\bibfield
  {title} {\enquote {\bibinfo {title} {Clausius {inequality} and {optimality}
  of {quasistatic} {transformations} for {nonequilibrium} {stationary}
  states},}\ }\href
  {http://journals.aps.org/prl/abstract/10.1103/PhysRevLett.110.020601}
  {\bibfield  {journal} {\bibinfo  {journal} {Phys. Rev. Lett.}\ }\textbf
  {\bibinfo {volume} {110}},\ \bibinfo {pages} {020601} (\bibinfo {year}
  {2013})}\BibitemShut {NoStop}%
\bibitem [{\citenamefont {Prados}\ \emph {et~al.}(2011)\citenamefont {Prados},
  \citenamefont {Lasanta},\ and\ \citenamefont {Hurtado}}]{prados11a}%
  \BibitemOpen
  \bibfield  {author} {\bibinfo {author} {\bibfnamefont {A.}~\bibnamefont
  {Prados}}, \bibinfo {author} {\bibfnamefont {A.}~\bibnamefont {Lasanta}}, \
  and\ \bibinfo {author} {\bibfnamefont {P.~I.}\ \bibnamefont {Hurtado}},\
  }\bibfield  {title} {\enquote {\bibinfo {title} {Large fluctuations in driven
  dissipative media},}\ }\href
  {http://journals.aps.org/prl/abstract/10.1103/PhysRevLett.107.140601}
  {\bibfield  {journal} {\bibinfo  {journal} {Phys. Rev. Lett.}\ }\textbf
  {\bibinfo {volume} {107}},\ \bibinfo {pages} {140601} (\bibinfo {year}
  {2011})}\BibitemShut {NoStop}%
\bibitem [{\citenamefont {Prados}\ \emph {et~al.}(2012)\citenamefont {Prados},
  \citenamefont {Lasanta},\ and\ \citenamefont {Hurtado}}]{prados12a}%
  \BibitemOpen
  \bibfield  {author} {\bibinfo {author} {\bibfnamefont {A.}~\bibnamefont
  {Prados}}, \bibinfo {author} {\bibfnamefont {A.}~\bibnamefont {Lasanta}}, \
  and\ \bibinfo {author} {\bibfnamefont {P.~I.}\ \bibnamefont {Hurtado}},\
  }\bibfield  {title} {\enquote {\bibinfo {title} {Nonlinear driven diffusive
  systems with dissipation: {Fluctuating} hydrodynamics},}\ }\href
  {http://journals.aps.org/pre/abstract/10.1103/PhysRevE.86.031134} {\bibfield
  {journal} {\bibinfo  {journal} {Phys. Rev. E}\ }\textbf {\bibinfo {volume}
  {86}},\ \bibinfo {pages} {031134} (\bibinfo {year} {2012})}\BibitemShut
  {NoStop}%
\bibitem [{\citenamefont {Hurtado}\ \emph {et~al.}(2013)\citenamefont
  {Hurtado}, \citenamefont {Lasanta},\ and\ \citenamefont
  {Prados}}]{hurtado13a}%
  \BibitemOpen
  \bibfield  {author} {\bibinfo {author} {\bibfnamefont {P.~I.}\ \bibnamefont
  {Hurtado}}, \bibinfo {author} {\bibfnamefont {A.}~\bibnamefont {Lasanta}}, \
  and\ \bibinfo {author} {\bibfnamefont {A.}~\bibnamefont {Prados}},\
  }\bibfield  {title} {\enquote {\bibinfo {title} {Typical and rare
  fluctuations in nonlinear driven diffusive systems with dissipation},}\
  }\href {http://journals.aps.org/pre/abstract/10.1103/PhysRevE.88.022110}
  {\bibfield  {journal} {\bibinfo  {journal} {Phys. Rev. E}\ }\textbf {\bibinfo
  {volume} {88}},\ \bibinfo {pages} {022110} (\bibinfo {year}
  {2013})}\BibitemShut {NoStop}%
\bibitem [{\citenamefont {Jack}\ \emph {et~al.}(2015)\citenamefont {Jack},
  \citenamefont {Thompson},\ and\ \citenamefont {Sollich}}]{jack15a}%
  \BibitemOpen
  \bibfield  {author} {\bibinfo {author} {\bibfnamefont {R.~L.}\ \bibnamefont
  {Jack}}, \bibinfo {author} {\bibfnamefont {I.~R.}\ \bibnamefont {Thompson}},
  \ and\ \bibinfo {author} {\bibfnamefont {P.}~\bibnamefont {Sollich}},\
  }\bibfield  {title} {\enquote {\bibinfo {title} {Hyperuniformity and phase
  separation in biased ensembles of trajectories for diffusive systems},}\
  }\href {http://journals.aps.org/prl/abstract/10.1103/PhysRevLett.114.060601}
  {\bibfield  {journal} {\bibinfo  {journal} {Phys. Rev. Lett.}\ }\textbf
  {\bibinfo {volume} {114}},\ \bibinfo {pages} {060601} (\bibinfo {year}
  {2015})}\BibitemShut {NoStop}%
\bibitem [{\citenamefont {Bodineau}\ and\ \citenamefont
  {Lagouge}(2010)}]{bodineau10a}%
  \BibitemOpen
  \bibfield  {author} {\bibinfo {author} {\bibfnamefont {T.}~\bibnamefont
  {Bodineau}}\ and\ \bibinfo {author} {\bibfnamefont {M.}~\bibnamefont
  {Lagouge}},\ }\bibfield  {title} {\enquote {\bibinfo {title} {Current {large}
  {deviations} in a {driven} {dissipative} model},}\ }\href
  {http://link.springer.com/article/10.1007%2Fs10955-010-9934-7} {\bibfield
  {journal} {\bibinfo  {journal} {J. Stat. Phys.}\ }\textbf {\bibinfo {volume}
  {139}},\ \bibinfo {pages} {201--218} (\bibinfo {year} {2010})}\BibitemShut
  {NoStop}%
\bibitem [{\citenamefont {Krapivsky}\ \emph {et~al.}(2014)\citenamefont
  {Krapivsky}, \citenamefont {Mallick},\ and\ \citenamefont
  {Sadhu}}]{krapivsky14a}%
  \BibitemOpen
  \bibfield  {author} {\bibinfo {author} {\bibfnamefont {P.~L.}\ \bibnamefont
  {Krapivsky}}, \bibinfo {author} {\bibfnamefont {K.}~\bibnamefont {Mallick}},
  \ and\ \bibinfo {author} {\bibfnamefont {T.}~\bibnamefont {Sadhu}},\
  }\bibfield  {title} {\enquote {\bibinfo {title} {Large {deviations} in
  {single-file} diffusion},}\ }\href
  {http://journals.aps.org/prl/abstract/10.1103/PhysRevLett.113.078101}
  {\bibfield  {journal} {\bibinfo  {journal} {Phys. Rev. Lett.}\ }\textbf
  {\bibinfo {volume} {113}},\ \bibinfo {pages} {078101} (\bibinfo {year}
  {2014})}\BibitemShut {NoStop}%
\bibitem [{\citenamefont {Meerson}\ and\ \citenamefont
  {Sasorov}(2014)}]{meerson14a}%
  \BibitemOpen
  \bibfield  {author} {\bibinfo {author} {\bibfnamefont {B.}~\bibnamefont
  {Meerson}}\ and\ \bibinfo {author} {\bibfnamefont {P.~V.}\ \bibnamefont
  {Sasorov}},\ }\bibfield  {title} {\enquote {\bibinfo {title} {Extreme current
  fluctuations in lattice gases: {Beyond} nonequilibrium steady states},}\
  }\href {http://journals.aps.org/pre/abstract/10.1103/PhysRevE.89.010101}
  {\bibfield  {journal} {\bibinfo  {journal} {Phys. Rev. E}\ }\textbf {\bibinfo
  {volume} {89}},\ \bibinfo {pages} {010101} (\bibinfo {year}
  {2014})}\BibitemShut {NoStop}%
\bibitem [{\citenamefont {Bouchet}\ \emph {et~al.}(2016)\citenamefont
  {Bouchet}, \citenamefont {Gaw{\c e}dzki},\ and\ \citenamefont
  {Nardini}}]{bouchet16a}%
  \BibitemOpen
  \bibfield  {author} {\bibinfo {author} {\bibfnamefont {F.}~\bibnamefont
  {Bouchet}}, \bibinfo {author} {\bibfnamefont {K.}~\bibnamefont {Gaw{\c
  e}dzki}}, \ and\ \bibinfo {author} {\bibfnamefont {C.}~\bibnamefont
  {Nardini}},\ }\bibfield  {title} {\enquote {\bibinfo {title} {Perturbative
  calculation of quasi-potential in non-equilibrium diffusions: a mean-field
  example},}\ }\href
  {http://link.springer.com/article/10.1007%2Fs10955-016-1503-2} {\bibfield
  {journal} {\bibinfo  {journal} {J. Stat. Phys.}\ }\textbf {\bibinfo {volume}
  {163}},\ \bibinfo {pages} {1157--1210} (\bibinfo {year} {2016})}\BibitemShut
  {NoStop}%
\bibitem [{\citenamefont {Bodineau}\ and\ \citenamefont
  {Derrida}(2005)}]{bodineau05a}%
  \BibitemOpen
  \bibfield  {author} {\bibinfo {author} {\bibfnamefont {T.}~\bibnamefont
  {Bodineau}}\ and\ \bibinfo {author} {\bibfnamefont {B.}~\bibnamefont
  {Derrida}},\ }\bibfield  {title} {\enquote {\bibinfo {title} {Distribution of
  current in nonequilibrium diffusive systems and phase transitions},}\ }\href
  {http://journals.aps.org/pre/abstract/10.1103/PhysRevE.72.066110} {\bibfield
  {journal} {\bibinfo  {journal} {Phys. Rev. E}\ }\textbf {\bibinfo {volume}
  {72}},\ \bibinfo {pages} {066110} (\bibinfo {year} {2005})}\BibitemShut
  {NoStop}%
\bibitem [{\citenamefont {Hurtado}\ and\ \citenamefont
  {Garrido}(2011)}]{hurtado11a}%
  \BibitemOpen
  \bibfield  {author} {\bibinfo {author} {\bibfnamefont {P.~I.}\ \bibnamefont
  {Hurtado}}\ and\ \bibinfo {author} {\bibfnamefont {P.~L.}\ \bibnamefont
  {Garrido}},\ }\bibfield  {title} {\enquote {\bibinfo {title} {Spontaneous
  symmetry breaking at the fluctuating level},}\ }\href
  {http://journals.aps.org/prl/abstract/10.1103/PhysRevLett.107.180601}
  {\bibfield  {journal} {\bibinfo  {journal} {Phys. Rev. Lett.}\ }\textbf
  {\bibinfo {volume} {107}},\ \bibinfo {pages} {180601} (\bibinfo {year}
  {2011})}\BibitemShut {NoStop}%
\bibitem [{\citenamefont {P\'erez-Espigares}\ \emph {et~al.}(2013)\citenamefont
  {P\'erez-Espigares}, \citenamefont {Garrido},\ and\ \citenamefont
  {Hurtado}}]{perez-espigares13a}%
  \BibitemOpen
  \bibfield  {author} {\bibinfo {author} {\bibfnamefont {C.}~\bibnamefont
  {P\'erez-Espigares}}, \bibinfo {author} {\bibfnamefont {P.~L.}\ \bibnamefont
  {Garrido}}, \ and\ \bibinfo {author} {\bibfnamefont {P.~I.}\ \bibnamefont
  {Hurtado}},\ }\bibfield  {title} {\enquote {\bibinfo {title} {Dynamical phase
  transition for current statistics in a simple driven diffusive system},}\
  }\href {http://journals.aps.org/pre/abstract/10.1103/PhysRevE.87.032115}
  {\bibfield  {journal} {\bibinfo  {journal} {Phys. Rev. E}\ }\textbf {\bibinfo
  {volume} {87}},\ \bibinfo {pages} {032115} (\bibinfo {year}
  {2013})}\BibitemShut {NoStop}%
\bibitem [{\citenamefont {Zarfaty}\ and\ \citenamefont
  {Meerson}()}]{zarfaty16a}%
  \BibitemOpen
  \bibfield  {author} {\bibinfo {author} {\bibfnamefont {L.}~\bibnamefont
  {Zarfaty}}\ and\ \bibinfo {author} {\bibfnamefont {B.}~\bibnamefont
  {Meerson}},\ }\bibfield  {title} {\enquote {\bibinfo {title} {Statistics of
  large currents in the {Kipnis-Marchioro-Presutti} model in a ring
  geometry},}\ }\href
  {http://iopscience.iop.org/article/10.1088/1742-5468/2016/03/033304/meta}
  {\bibinfo  {journal} {J. Stat. Mech.: Theory Exp. (2016) P033304}\ }\BibitemShut {NoStop}%
\bibitem [{\citenamefont {Vaikuntanathan}\ \emph {et~al.}(2014)\citenamefont
  {Vaikuntanathan}, \citenamefont {Gingrich},\ and\ \citenamefont
  {Geissler}}]{vaikuntanathan14a}%
  \BibitemOpen
\bibfield  {journal} {  }\bibfield  {author} {\bibinfo {author} {\bibfnamefont
  {S.}~\bibnamefont {Vaikuntanathan}}, \bibinfo {author} {\bibfnamefont
  {T.~R.}\ \bibnamefont {Gingrich}}, \ and\ \bibinfo {author} {\bibfnamefont
  {P.~L.}\ \bibnamefont {Geissler}},\ }\bibfield  {title} {\enquote {\bibinfo
  {title} {Dynamic phase transitions in simple driven kinetic networks},}\
  }\href {http://journals.aps.org/pre/abstract/10.1103/PhysRevE.89.062108}
  {\bibfield  {journal} {\bibinfo  {journal} {Phys. Rev. E}\ }\textbf {\bibinfo
  {volume} {89}},\ \bibinfo {pages} {062108} (\bibinfo {year}
  {2014})}\BibitemShut {NoStop}%
\bibitem [{\citenamefont {Lam}\ \emph {et~al.}(2009)\citenamefont {Lam},
  \citenamefont {Kurchan},\ and\ \citenamefont {Levine}}]{lam09a}%
  \BibitemOpen
  \bibfield  {author} {\bibinfo {author} {\bibfnamefont {K.~D. N.~T.}\
  \bibnamefont {Lam}}, \bibinfo {author} {\bibfnamefont {J.}~\bibnamefont
  {Kurchan}}, \ and\ \bibinfo {author} {\bibfnamefont {D.}~\bibnamefont
  {Levine}},\ }\bibfield  {title} {\enquote {\bibinfo {title} {Order in
  extremal trajectories},}\ }\href
  {http://link.springer.com/article/10.1007/s10955-009-9828-8} {\bibfield
  {journal} {\bibinfo  {journal} {J. Stat. Phys.}\ }\textbf {\bibinfo {volume}
  {137}},\ \bibinfo {pages} {1079--1093} (\bibinfo {year} {2009})}\BibitemShut
  {NoStop}%
\bibitem [{\citenamefont {Chandler}\ and\ \citenamefont
  {Garrahan}(2010)}]{chandler10a}%
  \BibitemOpen
  \bibfield  {author} {\bibinfo {author} {\bibfnamefont {D.}~\bibnamefont
  {Chandler}}\ and\ \bibinfo {author} {\bibfnamefont {J.~P.}\ \bibnamefont
  {Garrahan}},\ }\bibfield  {title} {\enquote {\bibinfo {title} {Dynamics on
  the way to forming glass: bubbles in space-time.}}\ }\href
  {http://www.annualreviews.org/doi/abs/10.1146/annurev.physchem.040808.090405}
  {\bibfield  {journal} {\bibinfo  {journal} {Annu. Rev. Phys. Chem.}\ }\textbf
  {\bibinfo {volume} {61}},\ \bibinfo {pages} {191--217} (\bibinfo {year}
  {2010})}\BibitemShut {NoStop}%
\bibitem [{\citenamefont {Shpielberg}\ and\ \citenamefont
  {Akkermans}(2016)}]{shpielberg16a}%
  \BibitemOpen
  \bibfield  {author} {\bibinfo {author} {\bibfnamefont {O.}~\bibnamefont
  {Shpielberg}}\ and\ \bibinfo {author} {\bibfnamefont {E.}~\bibnamefont
  {Akkermans}},\ }\bibfield  {title} {\enquote {\bibinfo {title} {Le
  {C}hatelier principle for out-of-equilibrium and boundary-driven systems:
  Application to dynamical phase transitions},}\ }\href
  {http://dx.doi.org/10.1103/PhysRevLett.116.240603} {\bibfield  {journal}
  {\bibinfo  {journal} {Phys. Rev. Lett.}\ }\textbf {\bibinfo {volume} {116}},\ \bibinfo {pages} {240603}
  (\bibinfo {year} {2016})}\BibitemShut {NoStop}%
\bibitem [{\citenamefont {Baek}\ \emph {et~al.}(2016)\citenamefont {Baek},
  \citenamefont {Kafri},\ and\ \citenamefont {Lecomte}}]{baek16a}%
  \BibitemOpen
  \bibfield  {author} {\bibinfo {author} {\bibfnamefont {Y.}~\bibnamefont
  {Baek}}, \bibinfo {author} {\bibfnamefont {Y.}~\bibnamefont {Kafri}}, \ and\
  \bibinfo {author} {\bibfnamefont {V.}~\bibnamefont {Lecomte}},\ }\bibfield
  {title} {\enquote {\bibinfo {title} {Dynamical symmetry breaking and phase
  transitions in driven diffusive systems},}\ }\href
  {http://journals.aps.org/prl/abstract/10.1103/PhysRevLett.118.030604} {\bibfield  {journal} {\bibinfo  {journal}
  {Phys. Rev. Lett.}\ }\textbf {\bibinfo {volume} {118}},\ \bibinfo {pages} {030604}  (\bibinfo {year} {2017})}\BibitemShut {NoStop}%
\bibitem [{\citenamefont {Tiz{\'o}n-Escamilla}\ \emph
  {et~al.}(2016)\citenamefont {Tiz{\'o}n-Escamilla}, \citenamefont
  {P{\'e}rez-Espigares}, \citenamefont {Garrido},\ and\ \citenamefont
  {Hurtado}}]{tizon-escamilla16a}%
  \BibitemOpen
  \bibfield  {author} {\bibinfo {author} {\bibfnamefont {N.}~\bibnamefont
  {Tiz{\'o}n-Escamilla}}, \bibinfo {author} {\bibfnamefont {C.}~\bibnamefont
  {P{\'e}rez-Espigares}}, \bibinfo {author} {\bibfnamefont {P.~L.}\
  \bibnamefont {Garrido}}, \ and\ \bibinfo {author} {\bibfnamefont {P.~I.}\
  \bibnamefont {Hurtado}},\ }\bibfield  {title} {\enquote {\bibinfo {title}
  {Order and symmetry-breaking in the fluctuations of driven systems},}\ }\href
  {https://arxiv.org/abs/1606.07507} {\bibfield  {journal} {\bibinfo  {journal}
  {arXiv:1606.07507}\ } (\bibinfo {year} {2016})}\BibitemShut {NoStop}%
\bibitem [{\citenamefont {Evans}\ \emph {et~al.}(1993)\citenamefont {Evans},
  \citenamefont {Cohen},\ and\ \citenamefont {Morriss}}]{evans93a}%
  \BibitemOpen
  \bibfield  {author} {\bibinfo {author} {\bibfnamefont {D.~J.}\ \bibnamefont
  {Evans}}, \bibinfo {author} {\bibfnamefont {E.~G.~D.}\ \bibnamefont {Cohen}},
  \ and\ \bibinfo {author} {\bibfnamefont {G.~P.}\ \bibnamefont {Morriss}},\
  }\bibfield  {title} {\enquote {\bibinfo {title} {Probability of 2nd law
  violations in shearing steady-states},}\ }\href
  {http://journals.aps.org/prl/abstract/10.1103/PhysRevLett.71.2401} {\bibfield
   {journal} {\bibinfo  {journal} {Phys. Rev. Lett.}\ }\textbf {\bibinfo
  {volume} {71}},\ \bibinfo {pages} {2401--2404} (\bibinfo {year}
  {1993})}\BibitemShut {NoStop}%
\bibitem [{\citenamefont {Gallavotti}\ and\ \citenamefont
  {Cohen}(1995{\natexlab{a}})}]{gallavotti95a}%
  \BibitemOpen
  \bibfield  {author} {\bibinfo {author} {\bibfnamefont {G.}~\bibnamefont
  {Gallavotti}}\ and\ \bibinfo {author} {\bibfnamefont {E.~G.~D.}\ \bibnamefont
  {Cohen}},\ }\bibfield  {title} {\enquote {\bibinfo {title} {Dynamical
  ensembles in nonequilibrium statistical-mechanics},}\ }\href
  {http://journals.aps.org/prl/abstract/10.1103/PhysRevLett.74.2694} {\bibfield
   {journal} {\bibinfo  {journal} {Phys. Rev. Lett.}\ }\textbf {\bibinfo
  {volume} {74}},\ \bibinfo {pages} {2694--2697} (\bibinfo {year}
  {1995}{\natexlab{a}})}\BibitemShut {NoStop}%
\bibitem [{\citenamefont {Gallavotti}\ and\ \citenamefont
  {Cohen}(1995{\natexlab{b}})}]{gallavotti95b}%
  \BibitemOpen
  \bibfield  {author} {\bibinfo {author} {\bibfnamefont {G.}~\bibnamefont
  {Gallavotti}}\ and\ \bibinfo {author} {\bibfnamefont {E.~G.~D.}\ \bibnamefont
  {Cohen}},\ }\bibfield  {title} {\enquote {\bibinfo {title} {Dynamical
  ensembles in stationary states},}\ }\href
  {http://link.springer.com/article/10.1007/BF02179860} {\bibfield  {journal}
  {\bibinfo  {journal} {J. Stat. Phys.}\ }\textbf {\bibinfo {volume} {80}},\
  \bibinfo {pages} {931--970} (\bibinfo {year}
  {1995}{\natexlab{b}})}\BibitemShut {NoStop}%
\bibitem [{\citenamefont {Kurchan}(1998)}]{kurchan98a}%
  \BibitemOpen
  \bibfield  {author} {\bibinfo {author} {\bibfnamefont {J.}~\bibnamefont
  {Kurchan}},\ }\bibfield  {title} {\enquote {\bibinfo {title} {Fluctuation
  theorem for stochastic dynamics},}\ }\href
  {http://iopscience.iop.org/article/10.1088/0305-4470/31/16/003/meta}
  {\bibfield  {journal} {\bibinfo  {journal} {J. Phys. A}\ }\textbf {\bibinfo
  {volume} {31}},\ \bibinfo {pages} {3719--3729} (\bibinfo {year}
  {1998})}\BibitemShut {NoStop}%
\bibitem [{\citenamefont {Lebowitz}\ and\ \citenamefont
  {Spohn}(1999)}]{lebowitz99a}%
  \BibitemOpen
  \bibfield  {author} {\bibinfo {author} {\bibfnamefont {J.~L.}\ \bibnamefont
  {Lebowitz}}\ and\ \bibinfo {author} {\bibfnamefont {H.}~\bibnamefont
  {Spohn}},\ }\bibfield  {title} {\enquote {\bibinfo {title} {A
  {Gallavotti-Cohen-type} symmetry in the large deviation functional for
  stochastic dynamics},}\ }\href
  {http://link.springer.com/article/10.1023%2FA%3A1004589714161} {\bibfield
  {journal} {\bibinfo  {journal} {J. Stat. Phys.}\ }\textbf {\bibinfo {volume}
  {95}},\ \bibinfo {pages} {333--365} (\bibinfo {year} {1999})}\BibitemShut
  {NoStop}%
\bibitem [{\citenamefont {Andrieux}\ and\ \citenamefont
  {Gaspard}()}]{andrieux07a}%
  \BibitemOpen
  \bibfield  {author} {\bibinfo {author} {\bibfnamefont {D.}~\bibnamefont
  {Andrieux}}\ and\ \bibinfo {author} {\bibfnamefont {P.}~\bibnamefont
  {Gaspard}},\ }\bibfield  {title} {\enquote {\bibinfo {title} {A fluctuation
  theorem for currents and non-linear response coefficients},}\ }\href
  {http://iopscience.iop.org/article/10.1088/1742-5468/2007/02/P02006/meta}
  {\bibinfo  {journal} {J. Stat. Mech.: Theory Exp. (2007) P02006}\ }\BibitemShut {NoStop}%
\bibitem [{\citenamefont {Gallavotti}(2014)}]{gallavotti14a}%
  \BibitemOpen
\bibfield  {journal} {  }\bibfield  {author} {\bibinfo {author} {\bibfnamefont
  {G.}~\bibnamefont {Gallavotti}},\ }\href@noop {} {\emph {\bibinfo {title}
  {Nonequilibrium and Irreversibility}}},\ Theoretical and Mathematical
  Physics\ (\bibinfo  {publisher} {Springer},\ \bibinfo {year}
  {2014})\BibitemShut {NoStop}%
\bibitem [{\citenamefont {Hurtado}\ \emph {et~al.}(2011)\citenamefont
  {Hurtado}, \citenamefont {P\'erez-Espigares}, \citenamefont {del Pozo},\ and\
  \citenamefont {Garrido}}]{hurtado11b}%
  \BibitemOpen
  \bibfield  {author} {\bibinfo {author} {\bibfnamefont {P.~I.}\ \bibnamefont
  {Hurtado}}, \bibinfo {author} {\bibfnamefont {C.}~\bibnamefont
  {P\'erez-Espigares}}, \bibinfo {author} {\bibfnamefont {J.~J.}\ \bibnamefont
  {del Pozo}}, \ and\ \bibinfo {author} {\bibfnamefont {P.~L.}\ \bibnamefont
  {Garrido}},\ }\bibfield  {title} {\enquote {\bibinfo {title} {Symmetries in
  fluctuations far from equilibrium},}\ }\href
  {http://www.pnas.org/content/108/19/7704.short} {\bibfield  {journal}
  {\bibinfo  {journal} {Proc. Natl. Acad. Sci. USA}\ }\textbf {\bibinfo
  {volume} {108}},\ \bibinfo {pages} {7704--7709} (\bibinfo {year}
  {2011})}\BibitemShut {NoStop}%
\bibitem [{\citenamefont {Villavicencio-Sanchez}\ \emph
  {et~al.}(2014)\citenamefont {Villavicencio-Sanchez}, \citenamefont {Harris},\
  and\ \citenamefont {Touchette}}]{villavicencio14a}%
  \BibitemOpen
  \bibfield  {author} {\bibinfo {author} {\bibfnamefont {R.}~\bibnamefont
  {Villavicencio-Sanchez}}, \bibinfo {author} {\bibfnamefont {R.~J.}\
  \bibnamefont {Harris}}, \ and\ \bibinfo {author} {\bibfnamefont
  {H.}~\bibnamefont {Touchette}},\ }\bibfield  {title} {\enquote {\bibinfo
  {title} {Fluctuation relations for anisotropic systems},}\ }\href
  {http://iopscience.iop.org/article/10.1209/0295-5075/105/30009/meta}
  {\bibfield  {journal} {\bibinfo  {journal} {Europhys. Lett.}\ }\textbf
  {\bibinfo {volume} {105}},\ \bibinfo {pages} {30009} (\bibinfo {year}
  {2014})}\BibitemShut {NoStop}%
\bibitem [{\citenamefont {Lacoste}\ and\ \citenamefont
  {Gaspard}(2014)}]{lacoste14a}%
  \BibitemOpen
  \bibfield  {author} {\bibinfo {author} {\bibfnamefont {D.}~\bibnamefont
  {Lacoste}}\ and\ \bibinfo {author} {\bibfnamefont {P.}~\bibnamefont
  {Gaspard}},\ }\bibfield  {title} {\enquote {\bibinfo {title} {Isometric
  {fluctuation} {relations} for {equilibrium} {states} with {broken}
  symmetry},}\ }\href
  {http://journals.aps.org/prl/abstract/10.1103/PhysRevLett.113.240602}
  {\bibfield  {journal} {\bibinfo  {journal} {Phys. Rev. Lett.}\ }\textbf
  {\bibinfo {volume} {113}},\ \bibinfo {pages} {240602} (\bibinfo {year}
  {2014})}\BibitemShut {NoStop}%
\bibitem [{\citenamefont {Gaspard}(2013)}]{gaspard13a}%
  \BibitemOpen
  \bibfield  {author} {\bibinfo {author} {\bibfnamefont {P.}~\bibnamefont
  {Gaspard}},\ }\bibfield  {title} {\enquote {\bibinfo {title} {Multivariate
  fluctuation relations for currents},}\ }\href
  {http://iopscience.iop.org/article/10.1088/1367-2630/15/11/115014/meta}
  {\bibfield  {journal} {\bibinfo  {journal} {New J. Phys.}\ }\textbf {\bibinfo
  {volume} {15}},\ \bibinfo {pages} {115014} (\bibinfo {year}
  {2013})}\BibitemShut {NoStop}%
\bibitem [{\citenamefont {P\'erez-Espigares}\ \emph {et~al.}(2015)\citenamefont
  {P\'erez-Espigares}, \citenamefont {Redig},\ and\ \citenamefont
  {Giardin\`a}}]{perez-espigares15a}%
  \BibitemOpen
  \bibfield  {author} {\bibinfo {author} {\bibfnamefont {C.}~\bibnamefont
  {P\'erez-Espigares}}, \bibinfo {author} {\bibfnamefont {F.}~\bibnamefont
  {Redig}}, \ and\ \bibinfo {author} {\bibfnamefont {C.}~\bibnamefont
  {Giardin\`a}},\ }\bibfield  {title} {\enquote {\bibinfo {title} {Spatial
  fluctuation theorem},}\ }\href
  {http://iopscience.iop.org/article/10.1088/1751-8113/48/35/35FT01/meta}
  {\bibfield  {journal} {\bibinfo  {journal} {J. Phys. A}\ }\textbf {\bibinfo
  {volume} {48}},\ \bibinfo {pages} {35FT01} (\bibinfo {year}
  {2015})}\BibitemShut {NoStop}%
\bibitem [{\citenamefont {Kumar}\ \emph {et~al.}(2015)\citenamefont {Kumar},
  \citenamefont {Soni}, \citenamefont {Ramaswamy},\ and\ \citenamefont
  {Sood}}]{kumar15a}%
  \BibitemOpen
  \bibfield  {author} {\bibinfo {author} {\bibfnamefont {N.}~\bibnamefont
  {Kumar}}, \bibinfo {author} {\bibfnamefont {H.}~\bibnamefont {Soni}},
  \bibinfo {author} {\bibfnamefont {S.}~\bibnamefont {Ramaswamy}}, \ and\
  \bibinfo {author} {\bibfnamefont {A.~K.}\ \bibnamefont {Sood}},\ }\bibfield
  {title} {\enquote {\bibinfo {title} {Anisotropic isometric fluctuation
  relations in experiment and theory on a self-propelled rod},}\ }\href
  {http://journals.aps.org/pre/abstract/10.1103/PhysRevE.91.030102} {\bibfield
  {journal} {\bibinfo  {journal} {Phys. Rev. E}\ }\textbf {\bibinfo {volume}
  {91}},\ \bibinfo {pages} {030102} (\bibinfo {year} {2015})}\BibitemShut
  {NoStop}%
\bibitem [{\citenamefont {Bodineau}\ and\ \citenamefont
  {Derrida}(2004)}]{bodineau04a}%
  \BibitemOpen
  \bibfield  {author} {\bibinfo {author} {\bibfnamefont {T.}~\bibnamefont
  {Bodineau}}\ and\ \bibinfo {author} {\bibfnamefont {B.}~\bibnamefont
  {Derrida}},\ }\bibfield  {title} {\enquote {\bibinfo {title} {Current
  fluctuations in nonequilibrium diffusive systems: {An} additivity
  principle},}\ }\href
  {http://journals.aps.org/prl/abstract/10.1103/PhysRevLett.92.180601}
  {\bibfield  {journal} {\bibinfo  {journal} {Phys. Rev. Lett.}\ }\textbf
  {\bibinfo {volume} {92}},\ \bibinfo {pages} {180601} (\bibinfo {year}
  {2004})}\BibitemShut {NoStop}%
\bibitem [{\citenamefont {Pilgram}\ \emph {et~al.}(2003)\citenamefont
  {Pilgram}, \citenamefont {Jordan}, \citenamefont {Sukhorukov},\ and\
  \citenamefont {B{\"u}ttiker}}]{pilgram03a}%
  \BibitemOpen
  \bibfield  {author} {\bibinfo {author} {\bibfnamefont {S.}~\bibnamefont
  {Pilgram}}, \bibinfo {author} {\bibfnamefont {A.~N.}\ \bibnamefont {Jordan}},
  \bibinfo {author} {\bibfnamefont {E.V.}\ \bibnamefont {Sukhorukov}}, \ and\
  \bibinfo {author} {\bibfnamefont {M.}~\bibnamefont {B{\"u}ttiker}},\
  }\bibfield  {title} {\enquote {\bibinfo {title} {Stochastic path integral
  formulation of full counting statistics},}\ }\href
  {http://journals.aps.org/prl/abstract/10.1103/PhysRevLett.90.206801}
  {\bibfield  {journal} {\bibinfo  {journal} {Phys. Rev. Lett.}\ }\textbf
  {\bibinfo {volume} {90}},\ \bibinfo {pages} {206801} (\bibinfo {year}
  {2003})}\BibitemShut {NoStop}%
\bibitem [{\citenamefont {Jordan}\ \emph {et~al.}(2004)\citenamefont {Jordan},
  \citenamefont {Sukhorukov},\ and\ \citenamefont {Pilgram}}]{jordan04a}%
  \BibitemOpen
  \bibfield  {author} {\bibinfo {author} {\bibfnamefont {A.~N.}\ \bibnamefont
  {Jordan}}, \bibinfo {author} {\bibfnamefont {E.~V.}\ \bibnamefont
  {Sukhorukov}}, \ and\ \bibinfo {author} {\bibfnamefont {S.}~\bibnamefont
  {Pilgram}},\ }\bibfield  {title} {\enquote {\bibinfo {title} {Fluctuation
  statistics in networks: A stochastic path integral approach},}\ }\href
  {http://scitation.aip.org/content/aip/journal/jmp/45/11/10.1063/1.1803927}
  {\bibfield  {journal} {\bibinfo  {journal} {J. Math. Phys.}\ }\textbf
  {\bibinfo {volume} {45}},\ \bibinfo {pages} {4386--4417} (\bibinfo {year}
  {2004})}\BibitemShut {NoStop}%
\bibitem [{\citenamefont {Bodineau}\ and\ \citenamefont
  {Derrida}(2006)}]{bodineau06a}%
  \BibitemOpen
  \bibfield  {author} {\bibinfo {author} {\bibfnamefont {T.}~\bibnamefont
  {Bodineau}}\ and\ \bibinfo {author} {\bibfnamefont {B.}~\bibnamefont
  {Derrida}},\ }\bibfield  {title} {\enquote {\bibinfo {title} {Current large
  deviations for asymmetric exclusion processes with open boundaries},}\ }\href
  {http://link.springer.com/article/10.1007/s10955-006-9048-4} {\bibfield
  {journal} {\bibinfo  {journal} {J. Stat. Phys.}\ }\textbf {\bibinfo {volume}
  {123}},\ \bibinfo {pages} {277--300} (\bibinfo {year} {2006})}\BibitemShut
  {NoStop}%
\bibitem [{\citenamefont {Hurtado}\ and\ \citenamefont
  {Garrido}(2009)}]{hurtado09c}%
  \BibitemOpen
  \bibfield  {author} {\bibinfo {author} {\bibfnamefont {P.~I.}\ \bibnamefont
  {Hurtado}}\ and\ \bibinfo {author} {\bibfnamefont {P.~L.}\ \bibnamefont
  {Garrido}},\ }\bibfield  {title} {\enquote {\bibinfo {title} {Test of the
  {additivity} {principle} for {current} {fluctuations} in a {model} of {heat}
  conduction},}\ }\href
  {http://journals.aps.org/prl/abstract/10.1103/PhysRevLett.102.250601}
  {\bibfield  {journal} {\bibinfo  {journal} {Phys. Rev. Lett.}\ }\textbf
  {\bibinfo {volume} {102}},\ \bibinfo {pages} {250601} (\bibinfo {year}
  {2009})}\BibitemShut {NoStop}%
\bibitem [{\citenamefont {Hurtado}\ and\ \citenamefont
  {Garrido}(2010)}]{hurtado10a}%
  \BibitemOpen
  \bibfield  {author} {\bibinfo {author} {\bibfnamefont {P.~I.}\ \bibnamefont
  {Hurtado}}\ and\ \bibinfo {author} {\bibfnamefont {P.~L.}\ \bibnamefont
  {Garrido}},\ }\bibfield  {title} {\enquote {\bibinfo {title} {Large
  fluctuations of the macroscopic current in diffusive systems: {A} numerical
  test of the additivity principle},}\ }\href
  {http://journals.aps.org/pre/abstract/10.1103/PhysRevE.81.041102} {\bibfield
  {journal} {\bibinfo  {journal} {Phys. Rev. E}\ }\textbf {\bibinfo {volume}
  {81}},\ \bibinfo {pages} {041102} (\bibinfo {year} {2010})}\BibitemShut
  {NoStop}%
\bibitem [{\citenamefont {Gorissen}\ and\ \citenamefont
  {Vanderzande}(2012)}]{gorissen12a}%
  \BibitemOpen
  \bibfield  {author} {\bibinfo {author} {\bibfnamefont {M.}~\bibnamefont
  {Gorissen}}\ and\ \bibinfo {author} {\bibfnamefont {C.}~\bibnamefont
  {Vanderzande}},\ }\bibfield  {title} {\enquote {\bibinfo {title} {Current
  fluctuations in the weakly asymmetric exclusion process with open
  boundaries},}\ }\href
  {http://journals.aps.org/pre/abstract/10.1103/PhysRevE.86.051114} {\bibfield
  {journal} {\bibinfo  {journal} {Phys. Rev. E}\ }\textbf {\bibinfo {volume}
  {86}},\ \bibinfo {pages} {051114} (\bibinfo {year} {2012})}\BibitemShut
  {NoStop}%
\bibitem [{\citenamefont {Gorissen}\ \emph {et~al.}(2012)\citenamefont
  {Gorissen}, \citenamefont {Lazarescu}, \citenamefont {Mallick},\ and\
  \citenamefont {Vanderzande}}]{gorissen12b}%
  \BibitemOpen
  \bibfield  {author} {\bibinfo {author} {\bibfnamefont {M.}~\bibnamefont
  {Gorissen}}, \bibinfo {author} {\bibfnamefont {A.}~\bibnamefont {Lazarescu}},
  \bibinfo {author} {\bibfnamefont {K.}~\bibnamefont {Mallick}}, \ and\
  \bibinfo {author} {\bibfnamefont {C.}~\bibnamefont {Vanderzande}},\
  }\bibfield  {title} {\enquote {\bibinfo {title} {Exact {current} {statistics}
  of the {asymmetric} {simple} {exclusion} {process} with {open} boundaries},}\
  }\href {http://journals.aps.org/prl/abstract/10.1103/PhysRevLett.109.170601}
  {\bibfield  {journal} {\bibinfo  {journal} {Phys. Rev. Lett.}\ }\textbf
  {\bibinfo {volume} {109}},\ \bibinfo {pages} {170601} (\bibinfo {year}
  {2012})}\BibitemShut {NoStop}%
\bibitem [{\citenamefont {Lazarescu}(2015)}]{lazarescu15a}%
  \BibitemOpen
  \bibfield  {author} {\bibinfo {author} {\bibfnamefont {A.}~\bibnamefont
  {Lazarescu}},\ }\bibfield  {title} {\enquote {\bibinfo {title} {The
  physicist's companion to current fluctuations: one-dimensional bulk-driven
  lattice gases},}\ }\href
  {http://iopscience.iop.org/article/10.1088/1751-8113/48/50/503001/meta}
  {\bibfield  {journal} {\bibinfo  {journal} {J. Phys. A}\ }\textbf {\bibinfo
  {volume} {48}},\ \bibinfo {pages} {503001} (\bibinfo {year}
  {2015})}\BibitemShut {NoStop}%
\bibitem [{\citenamefont {Akkermans}\ \emph {et~al.}(2013)\citenamefont
  {Akkermans}, \citenamefont {Bodineau}, \citenamefont {Derrida},\ and\
  \citenamefont {Shpielberg}}]{akkermans13a}%
  \BibitemOpen
  \bibfield  {author} {\bibinfo {author} {\bibfnamefont {E.}~\bibnamefont
  {Akkermans}}, \bibinfo {author} {\bibfnamefont {T.}~\bibnamefont {Bodineau}},
  \bibinfo {author} {\bibfnamefont {B.}~\bibnamefont {Derrida}}, \ and\
  \bibinfo {author} {\bibfnamefont {O.}~\bibnamefont {Shpielberg}},\ }\bibfield
   {title} {\enquote {\bibinfo {title} {Universal current fluctuations in the
  symmetric exclusion process and other diffusive systems},}\ }\href
  {http://iopscience.iop.org/article/10.1209/0295-5075/103/20001/meta}
  {\bibfield  {journal} {\bibinfo  {journal} {Europhys. Lett.}\ }\textbf
  {\bibinfo {volume} {103}},\ \bibinfo {pages} {20001} (\bibinfo {year}
  {2013})}\BibitemShut {NoStop}%
\bibitem [{\citenamefont {Becker}\ \emph {et~al.}(2015)\citenamefont {Becker},
  \citenamefont {Nelissen},\ and\ \citenamefont {Cleuren}}]{becker15a}%
  \BibitemOpen
  \bibfield  {author} {\bibinfo {author} {\bibfnamefont {T.}~\bibnamefont
  {Becker}}, \bibinfo {author} {\bibfnamefont {K.}~\bibnamefont {Nelissen}}, \
  and\ \bibinfo {author} {\bibfnamefont {B.}~\bibnamefont {Cleuren}},\
  }\bibfield  {title} {\enquote {\bibinfo {title} {Current fluctuations in
  boundary driven diffusive systems in different dimensions: a numerical
  study},}\ }\href
  {http://iopscience.iop.org/article/10.1088/1367-2630/17/5/055023/meta}
  {\bibfield  {journal} {\bibinfo  {journal} {New J. Phys.}\ }\textbf {\bibinfo
  {volume} {17}},\ \bibinfo {pages} {055023} (\bibinfo {year}
  {2015})}\BibitemShut {NoStop}%
\bibitem [{\citenamefont {P\'erez-Espigares}\ \emph {et~al.}(2016)\citenamefont
  {P\'erez-Espigares}, \citenamefont {Garrido},\ and\ \citenamefont
  {Hurtado}}]{perez-espigares16a}%
  \BibitemOpen
  \bibfield  {author} {\bibinfo {author} {\bibfnamefont {C.}~\bibnamefont
  {P\'erez-Espigares}}, \bibinfo {author} {\bibfnamefont {P.~L.}\ \bibnamefont
  {Garrido}}, \ and\ \bibinfo {author} {\bibfnamefont {P.~I.}\ \bibnamefont
  {Hurtado}},\ }\bibfield  {title} {\enquote {\bibinfo {title} {Weak additivity
  principle for current statistics in $d$-dimensions},}\ }\href
  {http://journals.aps.org/pre/abstract/10.1103/PhysRevE.93.040103} {\bibfield
  {journal} {\bibinfo  {journal} {Phys. Rev. E}\ }\textbf {\bibinfo {volume}
  {93}},\ \bibinfo {pages} {040103(R)} (\bibinfo {year} {2016})}\BibitemShut
  {NoStop}%
\bibitem [{\citenamefont {Villavicencio-Sanchez}\ and\ \citenamefont
  {Harris}(2016)}]{villavicencio16a}%
  \BibitemOpen
  \bibfield  {author} {\bibinfo {author} {\bibfnamefont {R.}~\bibnamefont
  {Villavicencio-Sanchez}}\ and\ \bibinfo {author} {\bibfnamefont {R.~J.}\
  \bibnamefont {Harris}},\ }\bibfield  {title} {\enquote {\bibinfo {title}
  {Local structure of current fluctuations in diffusive systems beyond one
  dimension},}\ }\href
  {http://journals.aps.org/pre/abstract/10.1103/PhysRevE.93.032134} {\bibfield
  {journal} {\bibinfo  {journal} {Phys. Rev. E}\ }\textbf {\bibinfo {volume}
  {93}},\ \bibinfo {pages} {032134} (\bibinfo {year} {2016})}\BibitemShut
  {NoStop}%
\bibitem [{\citenamefont {Martin}, \citenamefont {Siggia}\ and\ \citenamefont {Rose}(1973)}]{martin16a}%
  \BibitemOpen
  \bibfield  {author} {\bibinfo {author} {\bibfnamefont {P.C.}~\bibnamefont
  {Martin}}, \bibinfo {author} {\bibfnamefont {E.D.}\ \bibnamefont
  {Siggia}}\ and\ \bibinfo {author} {\bibfnamefont {H.A.}~\bibnamefont {Rose}},\
  }\bibfield  {title} {\enquote {\bibinfo {title} {Statistical dynamics of classical systems},}\ }\href
  {http://http://journals.aps.org/pra/pdf/10.1103/PhysRevA.8.423}
  {\bibfield  {journal} {\bibinfo  {journal} {Phys. Rev. A}\ }\textbf
  {\bibinfo {volume} {8}},\ \bibinfo {pages} {423} (\bibinfo {year}
  {1973})}\BibitemShut {NoStop}%
\bibitem [{\citenamefont {Arfken}\ \emph {et~al.}(2011)\citenamefont {Arfken},
  \citenamefont {Weber},\ and\ \citenamefont {Harris}}]{arfken11a}%
  \BibitemOpen
  \bibfield  {author} {\bibinfo {author} {\bibfnamefont {G.~B.}\ \bibnamefont
  {Arfken}}, \bibinfo {author} {\bibfnamefont {H.~J.}\ \bibnamefont {Weber}}, \
  and\ \bibinfo {author} {\bibfnamefont {F.~E.}\ \bibnamefont {Harris}},\
  }\href@noop {} {\emph {\bibinfo {title} {Mathematical methods for physicists:
  a comprehensive guide}}}\ (\bibinfo  {publisher} {Academic Press, San Diego},\ \bibinfo
  {year} {2011})\BibitemShut {NoStop}%
\bibitem [{\citenamefont {Saito}\ and\ \citenamefont {Dhar}(2011)}]{saito11a}%
  \BibitemOpen
  \bibfield  {author} {\bibinfo {author} {\bibfnamefont {K.}~\bibnamefont
  {Saito}}\ and\ \bibinfo {author} {\bibfnamefont {A.}~\bibnamefont {Dhar}},\
  }\bibfield  {title} {\enquote {\bibinfo {title} {Additivity {principle} in
  {high-dimensional} {deterministic} systems},}\ }\href
  {http://journals.aps.org/prl/abstract/10.1103/PhysRevLett.107.250601}
  {\bibfield  {journal} {\bibinfo  {journal} {Phys. Rev. Lett.}\ }\textbf
  {\bibinfo {volume} {107}},\ \bibinfo {pages} {250601} (\bibinfo {year}
  {2011})}\BibitemShut {NoStop}%
\end{thebibliography}
\end{document}